\documentclass[aps,prl,reprint,floatfix,superscriptaddress,nobibnotes]{revtex4-1}

\usepackage{amsmath}
\usepackage{amssymb}
\usepackage{graphicx}
\usepackage{hyperref}
\usepackage{color}
\usepackage[]{siunitx}
\hypersetup{
    colorlinks,
    linkcolor={blue},
    citecolor={blue},
    urlcolor={blue}
}

\begin{document}

\title{Supercurrent transport through 1$e$-periodic full-shell Coulomb islands}

\author{D.~Razmadze}
\altaffiliation{These authors contributed equally to this work.}
\affiliation{Center for Quantum Devices, Niels Bohr Institute, University of Copenhagen, 2100 Copenhagen, Denmark}
\author{R.~Seoane~Souto}
\altaffiliation{These authors contributed equally to this work.}
\affiliation{Center for Quantum Devices, Niels Bohr Institute, University of Copenhagen, 2100 Copenhagen, Denmark}
\affiliation{Division of Solid State Physics and NanoLund, Lund University, 22100 Lund, Sweden}
\affiliation{Instituto de Ciencia de Materiales de Madrid (ICMM), Consejo Superior de Investigaciones Científicas (CSIC),
Sor Juana Inés de la Cruz 3, 28049 Madrid, Spain}
\author{E.~C.~T.~O'Farrell}
\affiliation{Center for Quantum Devices, Niels Bohr Institute, University of Copenhagen, 2100 Copenhagen, Denmark}
\author{P.~Krogstrup}
\affiliation{Center for Quantum Devices, Niels Bohr Institute, University of Copenhagen, 2100 Copenhagen, Denmark}
\author{M.~Leijnse}
\affiliation{Center for Quantum Devices, Niels Bohr Institute, University of Copenhagen, 2100 Copenhagen, Denmark}
\affiliation{Division of Solid State Physics and NanoLund, Lund University, 22100 Lund, Sweden}
\author{C.~M.~Marcus}
\affiliation{Center for Quantum Devices, Niels Bohr Institute, University of Copenhagen, 2100 Copenhagen, Denmark}
\author{S.~Vaitiek\.{e}nas}
\affiliation{Center for Quantum Devices, Niels Bohr Institute, University of Copenhagen, 2100 Copenhagen, Denmark}

\date{\today}

\begin{abstract}
We experimentally investigate supercurrent through Coulomb islands, where island and leads are fabricated from semiconducting nanowires with fully surrounding superconducting shells.
Applying flux along the wire yields a series of destructive Little-Parks lobes with reentrant supercurrent.
We find Coulomb blockade with 2$e$ peak spacing in the zeroth lobe and 1$e$ average spacing, with regions of significant even-odd modulation, in the first lobe.
Evolution of Coulomb-peak amplitude through the first lobe is consistent with a theoretical model of supercurrent carried predominantly by zero-energy states in the leads and the island.
\end{abstract}

\maketitle

The interplay between the superconducting gap, $\Delta$, and charging energy, $E_{\rm C}$, in mesoscopic superconductors gives rise to intriguing phenomena sensitive to single electron charging events~\cite{Averin1992}.
Early transport studies revealed that the parity of a superconducting island depends sensitively on the ratio between $\Delta$ and $E_{\rm C}$~\cite{Tuominen1992, LaFarge1993, Lafarge1993_2}.
In particular, structures with $\Delta > E_{\rm C}$ can only be charged in multiples of 2$e$.
Suppressing $\Delta$ by temperature or magnetic field allows 1$e$ charging.
Thanks to their unique characteristics, superconducting islands are used as the basis of various modern quantum devices, including Cooper-pair transistors \cite{Ferguson2006, Pekola2008, vanWoerkom2015}, charge qubits \cite{Nakamura1999, Koch2007}, and sensitive photon detectors~\cite{Mannila2021, Tanarom2022}.

Subgap states, emerging in superconducting hybrids, introduce an additional energy scale that characterizes the transport properties of the systems~\cite{Pillet2010, Chang2013, Lee2014, Junger2020, Prada2020}.
In semiconductor-superconductor islands, the subgap-state energy, $\varepsilon$, can be tuned via gate voltage or spin-splitting fields, allowing for precise control of island parity~\cite{Higginbotham2015, Veen_PRB2018, Vaitiekenas2022}.
Islands with states at $\varepsilon < E_{\rm C}$ can host an odd number of electrons resulting in even-odd Coulomb blockade periodicity~\cite{Albrecht2017, Shen2018}; for $\varepsilon= 0$, the Coulomb blockade becomes 1$e$ periodic, which has been studied in the context of trivial~\cite{Valentini2022, Saldana2023} and topological~\cite{VanHeck2016, Albrecht2016, Vaitiekenas2020, Shen2021, Souto2022} superconductivity.
Previous experiments on devices with superconducting leads showed supercurrent in the Coulomb blockade regime~\cite{Ofarrell2018, Wang2022} and $4\pi$-periodic current-phase relation~\cite{Bocquillon2017, Laroche2019, Dartiailh2021}, theoretically attributed to the interplay between the Josephson coupling and low-energy subgap states~\cite{Jiang2011, Dominguez2017, Vuik2019, Souto2022_2}.

\begin{figure}[t!]
\includegraphics[width=\linewidth]{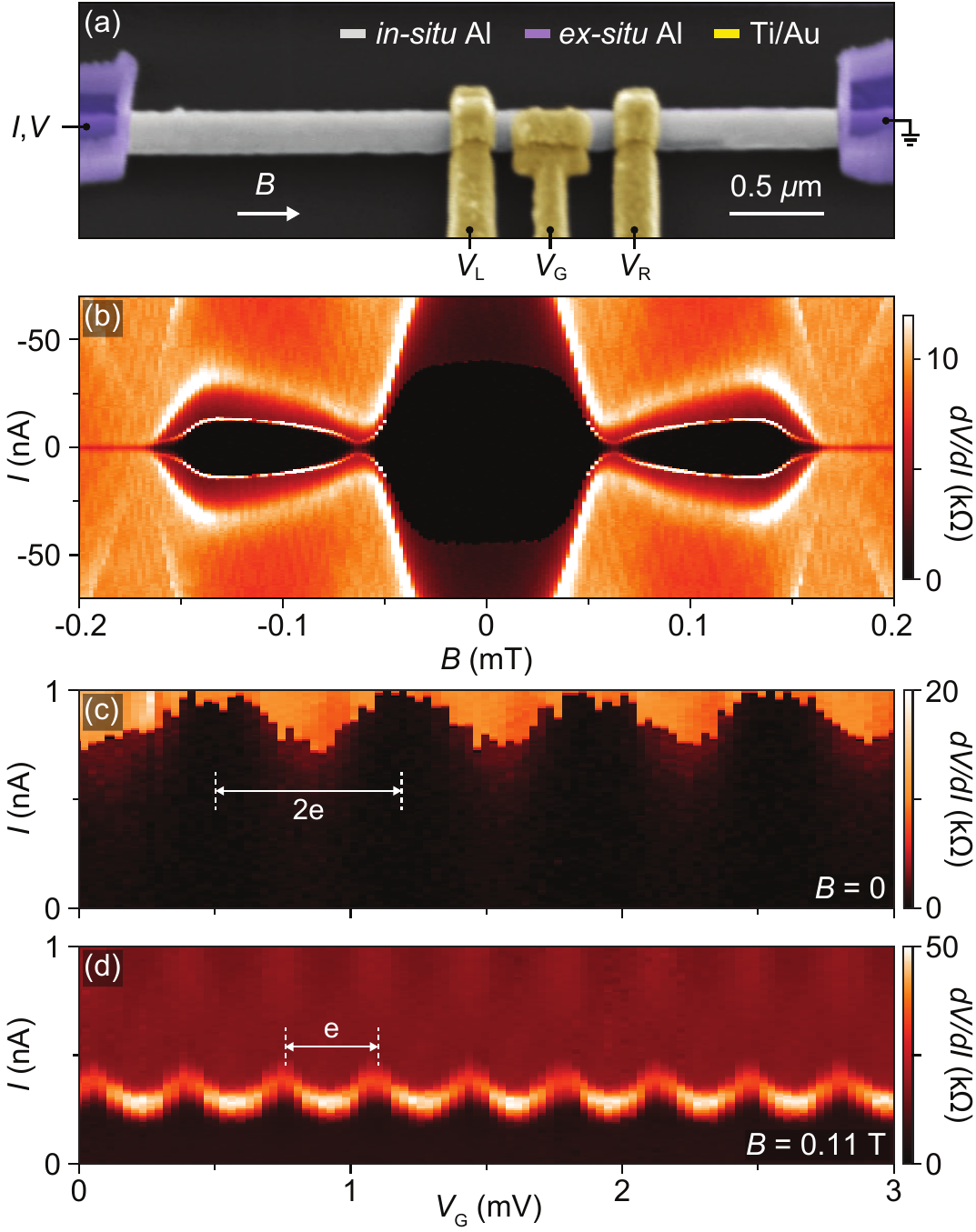}
\caption{\label{fig:1}
(a) Color-enhanced electron micrograph of device~1 comprising hybrid InAs-Al full-shell Coulomb island.
(b)~Differential resistance, $dV/dI$, as a function of current bias, $I$, and axial magnetic field, $B$, displaying periodic Little-Parks modulation of the switching current as the magnetic flux threads the wire. The data were taken in the open regime ($V_{\rm L}$ = $V_{\rm R}$ = 2~V).
(c) Zero-field $dV/dI$ as a function of $I$ measured in Coulomb blockade regime ($V_{\rm L}$ = 55~mV, $V_{\rm R}$ = -182~mV), showing 2$e$-periodic switching current modulation as a function of island-gate voltage, $V_{\rm G}$.
(d) Same as (c) but measured at $B=110$~mT, equivalent to one flux quantum threading the wire, showing roughly 1$e$-periodic switching current modulation by $V_{\rm G}$.
}
\end{figure}

Here, we use Coulomb-blockade transport to study the enhancement of 1$e$-periodic supercurrent in hybrid islands defined in semiconductor-superconductor full-shell nanowires.
A representative device is shown in Fig.~\ref{eq:1}(a).
Applying magnetic flux along the wire induces the destructive Little-Parks effect~\cite{Little1962}, resulting in a characteristic sequence of reentrant supercurrent lobes, as previously observed in cylindrical (or roughly cylindrical) samples with diameters below the superconducting coherence length~\cite{Liu2001, Sternfeld2011, Vaitiekenas2020, Vaitiekenas2020_2}.
In the zeroth lobe, transport displays 2$e$-periodic features in the Coulomb blockaded regime.
In contrast, the Coulomb blockade is roughly 1$e$ periodic throughout the first lobe, with small even-odd modulations, displaying an overall conductance background enhancement in the middle of the lobe.
By comparing the experiment to a theoretical model, we find that the measurements are consistent with the presence of low-energy states in all three device segments.

The measured devices consist of hexagonal InAs nanowires with epitaxial \textit{in situ} Al grown on all six facets, forming a continuous, fully surrounding superconducting shell~\cite{Krogstrup2015, Vaitiekenas2020, Kringhoj2021, Ibabe2023}.
Two $\sim$100~nm segments of the Al shell were selectively etched to form gateable Josephson junctions, controlled by gate voltages $V_{\rm L}$ and $V_{\rm R}$, see Fig.~\ref{fig:1}(a).
The charge offset of the island between the two junctions was tuned by the gate voltage $V_{\rm G}$.
The wire ends were contacted with Al leads, deposited in a separate lithography step.
The evolution of the current-phase relation with axial flux through these wires was recently investigated through a combination of interferometry experiments~\cite{Razmadze2020} and theoretical analysis~\cite{Schulenborg2020, Svetogorov2023}.
We report Coulomb spectroscopy results from four devices.
In the main text, we focus on data from device 1, comprising a 700~nm island.
Supporting data from the other three devices, including one with a 300~nm island, are presented in the Supplemental Material~\cite{Supplement}.
Transport measurements were carried out using standard ac lock-in techniques in a dilution refrigerator with a three-axis vector magnet and base temperature of 20 mK.

Differential resistance, $dV/dI$, as a function of current bias, $I$, and axial magnetic field, $B$, in the open regime ($V_{\rm L}$ = $V_{\rm R}$ = 2~V) reveals a series of superconducting lobes, see Fig.~\ref{fig:1}(b).
The switching current is maximal in the zeroth lobe, reaching roughly 50~nA at $B=0$.
The supercurrent vanishes around $\vert B \vert\sim55$~mT, corresponding to half flux quantum through the nanowire, but revives again at higher field values.
In the first lobe, the maximal switching current of $\sim 15$~nA is skewed toward the high-field end of the lobe.
This is likely because the bound states carrying the supercurrent through the junctions have a smaller cross-section compared to the parent Al shell and hence require a higher field to reach one flux quantum~\cite{Kopasov2020, Kringhoj2021, SanJose2023}.
The two resistance steps in the zeroth and first lobes presumably result from different switching currents in two junctions.
The sharp features, visible around $\vert B \vert \sim 170$~mT, can be associated with the switching current of the shell.

\begin{figure}[t!]
\includegraphics[width=\linewidth]{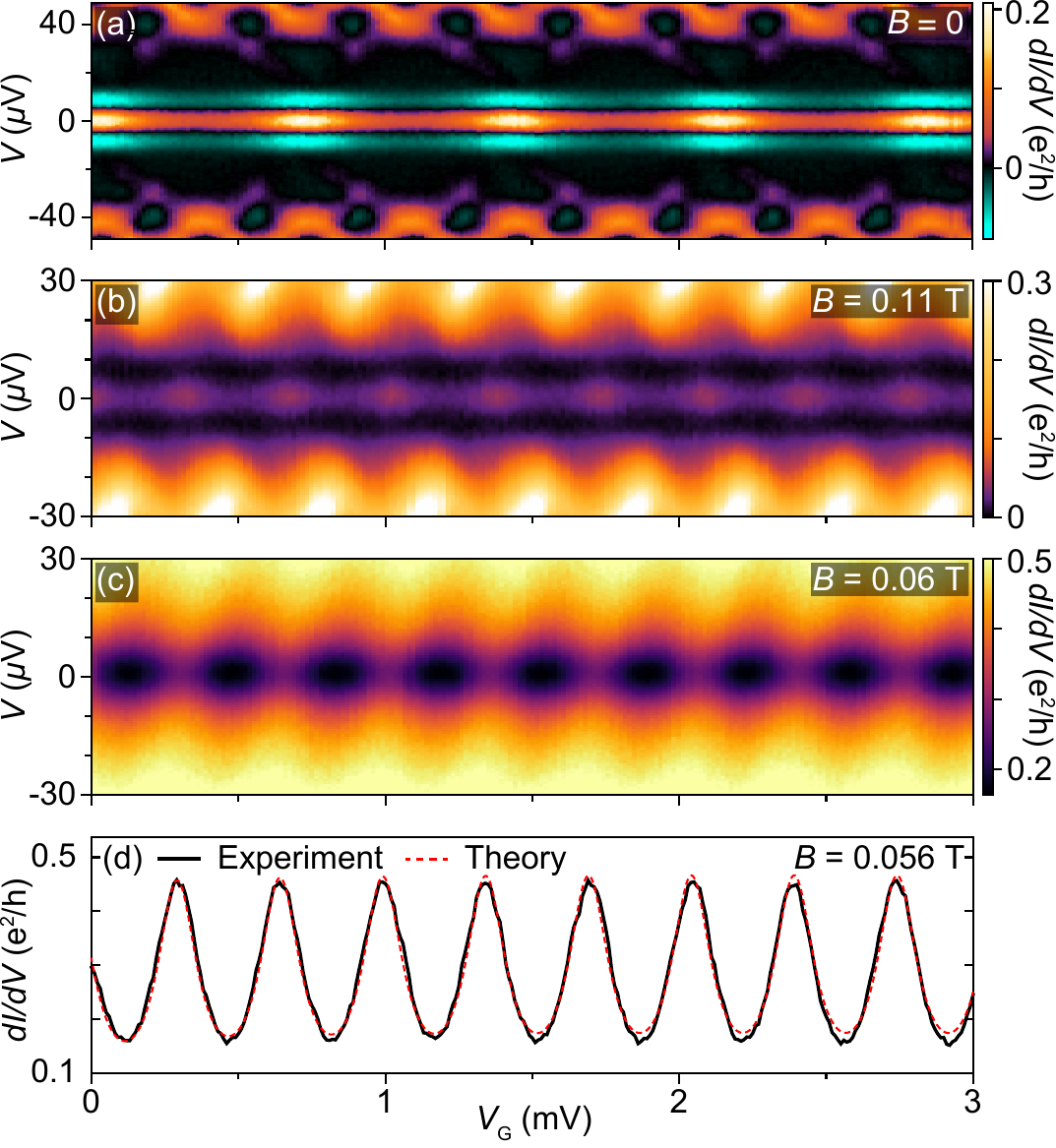}
\caption{\label{fig:2}
Differential conductance, $dI/dV$, as a function of source-drain voltage bias, $V$, and island-gate voltage, $V_{\rm G}$, in Coulomb-blockade regime ($V_{\rm L}$ = -0.17~V and $V_{\rm R}$ = -0.3~V), measured in (a) the zeroth lobe at $B=0$, (b) the first lobe at $B$ = 110~mT, and (c) the destructive regime at $B$ = 60~mT.
(d)~Zero-bias conductance measured in the destructive regime at $V_{\rm L}$ = -0.145~V and $V_{\rm R}$ = -0.45~V (black curve), compared to a theoretical model (red-dashed curve) giving effective temperature of $T\sim80$~mK and tunnel coupling to the two leads of $\hbar\Gamma_{\rm L}\sim\hbar\Gamma_{\rm R}/7 \sim 10$~$\mu$eV, see Ref.~\cite{Supplement}.
}
\end{figure}

Lowering the junction-gate voltages decreases the Josephson coupling and increases the charging energy of the island, tuning the device into the Coulomb blockade regime.
In this regime, the switching current oscillates as a function of $V_{\rm G}$ with a period of $\sim0.7$~mV around zero field [Fig.~\ref{fig:1}(c)] and roughly half of that at $B=110$~mT [Fig.~\ref{fig:1}(d)].
The factor-of-two decrease in gate-voltage period suggests that $\Delta$ in the zeroth lobe exceeds $E_{\rm C}$ of the island, allowing for 2$e$ charging.
In the first lobe, 1$e$ charging becomes possible due to decreased spectral gap.
The evenly spaced switching-current features at one flux quantum indicate the presence of subgap states, either continuous or discrete, near zero energy in the island.

To investigate the subgap excitations, we turn to Coulomb spectroscopy in a voltage-bias configuration.
Differential conductance, $dI/dV$, as a function of voltage bias, $V$, and $V_{\rm G}$, measured at $B=0$ in the Coulomb blockade regime exhibits an extended supercurrent peak around $V=0$, see Fig.~\ref{fig:2}(a).
Peak height is modulated periodically by $V_{\rm G}$ but remains nonzero throughout the measured range.
Additional features with half the period in $V_{\rm G}$ at higher bias correspond to multiple Andreev reflections that excite the island~\cite{Brink1991, Hadley1998}.
The first lobe displays a supercurrent peak similar to that of the zeroth lobe but with a halved period for its height modulations [Fig.~\ref{fig:2}(b)], consistent with the current-bias measurements in Fig.~\ref{fig:1}.

In the destructive regime between lobes, where superconductivity is suppressed, the device behaves as a metallic island coupled to normal leads.
Transport through the system exhibits Coulomb diamonds without discrete features that extend continuously to zero bias, unlike in the superconducting lobes, see Fig.~\ref{fig:2}(c).
Transport in this regime is well understood and can be used to extract effective electron temperature, $T$, and tunnel rates between the island and the leads, $\Gamma_{\rm L}$ and $\Gamma_{\rm R}$~\cite{Thijssen2008}.
Comparison of the zero-bias conductance to an expanded theory model~\cite{Supplement}, yields $T\sim 80$~mK and $\hbar\Gamma_{\rm L}\sim\hbar\Gamma_{\rm R}/7 \sim 10$~$\mu$eV.

\begin{figure}
\includegraphics[width=\linewidth]{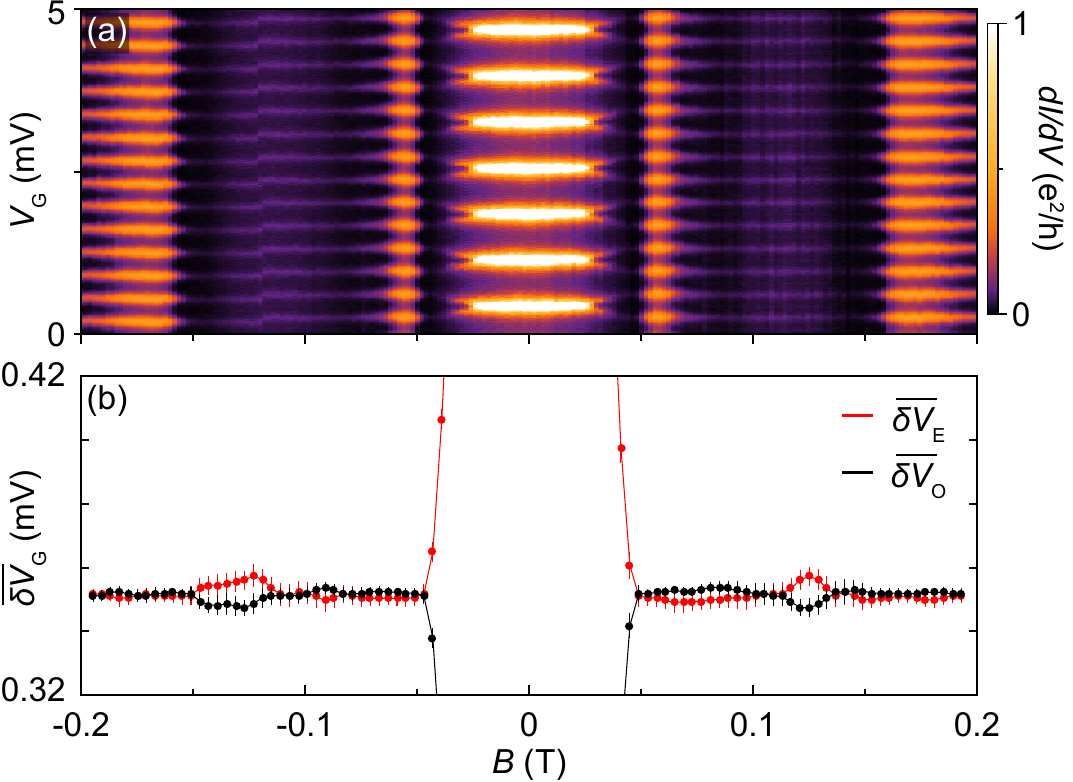}
\caption{\label{fig:3}
(a) Zero-bias conductance, $dI/dV$, measured in the weak coupling regime ($V_{\rm L}$ = -0.145~V and $V_{\rm R}$ = -0.45~V), showing Coulomb blockade evolution as a function of island-gate voltage, $V_{\rm G}$, and axial magnetic field, $B$.
The Coulomb peaks display a locally enhanced conductance around one flux quantum ($\vert B\vert\sim110$~mT).
(b) Average peak spacing, $\overline{\delta V}$, for even (red) and odd (black) Coulomb valleys, extracted from the data in (a).
Coulomb peaks are 2$e$-periodic in the middle of the zeroth lobe, but split around $\vert B\vert = 40$~mT and become 1$e$ periodic in the destructive regime ($\vert B\vert \sim 55$~mT).
The peaks remain roughly 1$e$ periodic with slight even-odd modulation throughout the first lobe.
}
\end{figure}

Focusing on zero-bias features, we study the evolution of Coulomb-peak periodicity with $B$ [Fig.~\ref{fig:3}(a)].
In the zeroth lobe, Coulomb peaks are 2$e$-periodic up to $\vert B \vert\sim40$~mT, then split as $\Delta$ decreases below $E_{\rm C}$.
A comparison with an independent gap measurement at the cross-over point yields $E_{\rm C}=120~\mu$eV~\cite{Supplement}.
For larger $B$, the peaks continue splitting, becoming 1$e$-periodic around $\vert B\vert = 55$~mT, at the onset of the destructive regime.
At yet higher field, in the first lobe, the peaks remain nearly 1$e$-periodic with slight even-odd modulation. We note that the conductance is finite for all gate-voltage values throughout the first lobe, consistent with the data in Fig.~\ref{fig:2}(b).
Deviation from 1$e$ periodicity is evident in the extracted average peak spacings, $\delta V_{\rm G}$, for even and odd Coulomb valleys, shown in Fig.~\ref{fig:3}(b).
The oscillatory behavior suggests a discrete subgap state crossing zero energy.

\begin{figure}[t!]
\includegraphics[width=\linewidth]{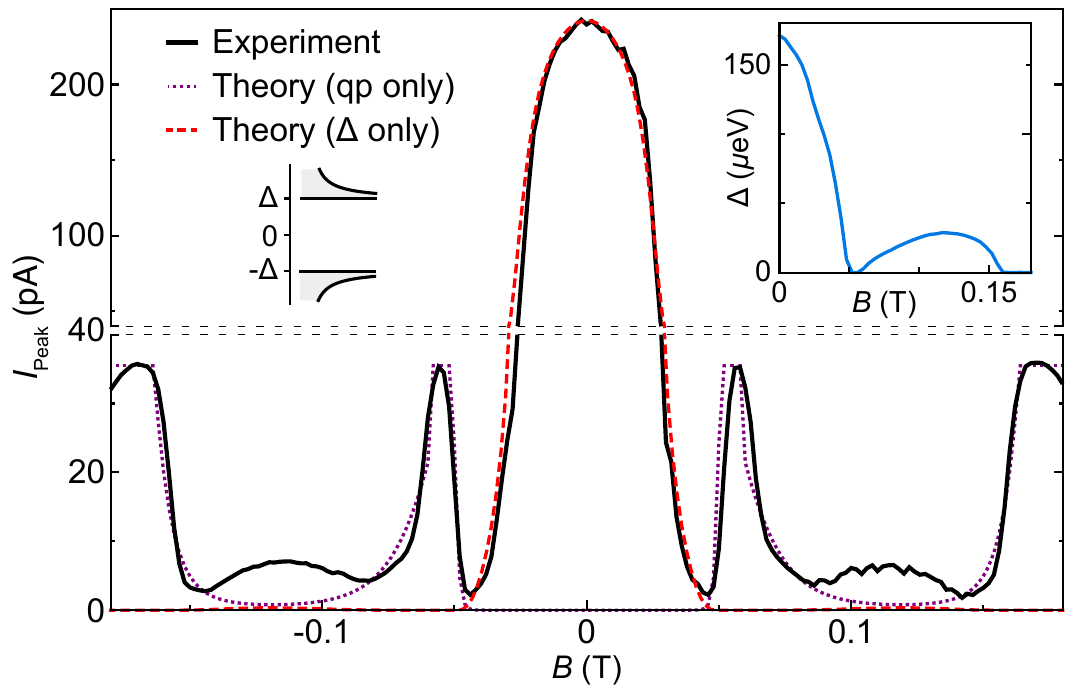}
\caption{\label{fig:4}
Average Coulomb-peak current, $I_{\rm Peak}$, as a function of axial magnetic field, $B$.
Experimental data (solid black curve) is extracted by integrating the differential conductance in Fig.~\ref{fig:3}(a) over the ac voltage excitation window.
Normal quasiparticle current (dotted purple curve), calculated using Eq.~\eqref{eq:2}, peaks in the destructive regimes and is exponentially suppressed in the superconducting lobes.
Theoretical supercurrent (dashed red curve), calculated using the zero-bandwidth model with contributions only from the continuum states, shows a good agreement with the data in the zeroth lobe, but does not explain the bump in the middle of the first lobe. 
Inset: Superconducting gap, $\Delta$, as a function of $B$, used as an input for both theory curves, was extracted from an independent tunneling spectroscopy measurement, see Ref.~\cite{Supplement}.
}
\end{figure}

Further analysis of peak spacings yield the lowest excitation energy~\cite{Vaitiekenas2018_2},
\begin{equation}\label{eq:1}
\varepsilon=E_{\rm C}\,\,\frac{\overline{\delta V}_{\rm E}-\overline{\delta V}_{\rm O}}{\overline{\delta V}_{\rm E}+\overline{\delta V}_{\rm O}},
\end{equation}
where $\overline{\delta V}_{\rm E(O)}$ is the ensemble-averaged gate-voltage spacing of even (odd) Coulomb peaks.
In the first lobe, the maximal even-odd peak spacing difference of $\sim10~\mu$V is consistent with a subgap state in the island with $\varepsilon=2~\mu$eV.
This value agrees with previously reported values for islands of similar length~\cite{Vaitiekenas2020}.
We note that device 4, with a shorter (300~nm) island, shows a larger even-odd peak spacing difference. When expressed in energy units, the difference corresponds roughly to the superconducting gap in the first lobe~\cite{Supplement}.

In addition to the even-odd peak spacing modulation in the first lobe, we observe a non-monotonic peak height dependence, displaying a maximum in the middle of the lobe, see Fig.~\ref{fig:3}(a).
This behavior is qualitatively consistent across different configurations of junction-gate voltages and for the other two devices with 700~nm islands, but absent in shorter 300~nm islands~\cite{Supplement} and devices with normal leads~\cite{Vaitiekenas2020}.
To better understand this feature, we develop a model of supercurrent flow through the island, considering electron pairing and charging energy on equal footing (see Supplemental Material~\cite{Supplement}).
The continuum states and the subgap states in the first lobe are described using the zero-bandwidth approximation~\cite{Affleck2000, Vecino2003, Grove2018}.

\begin{figure}[t!]
\includegraphics[width=\linewidth]{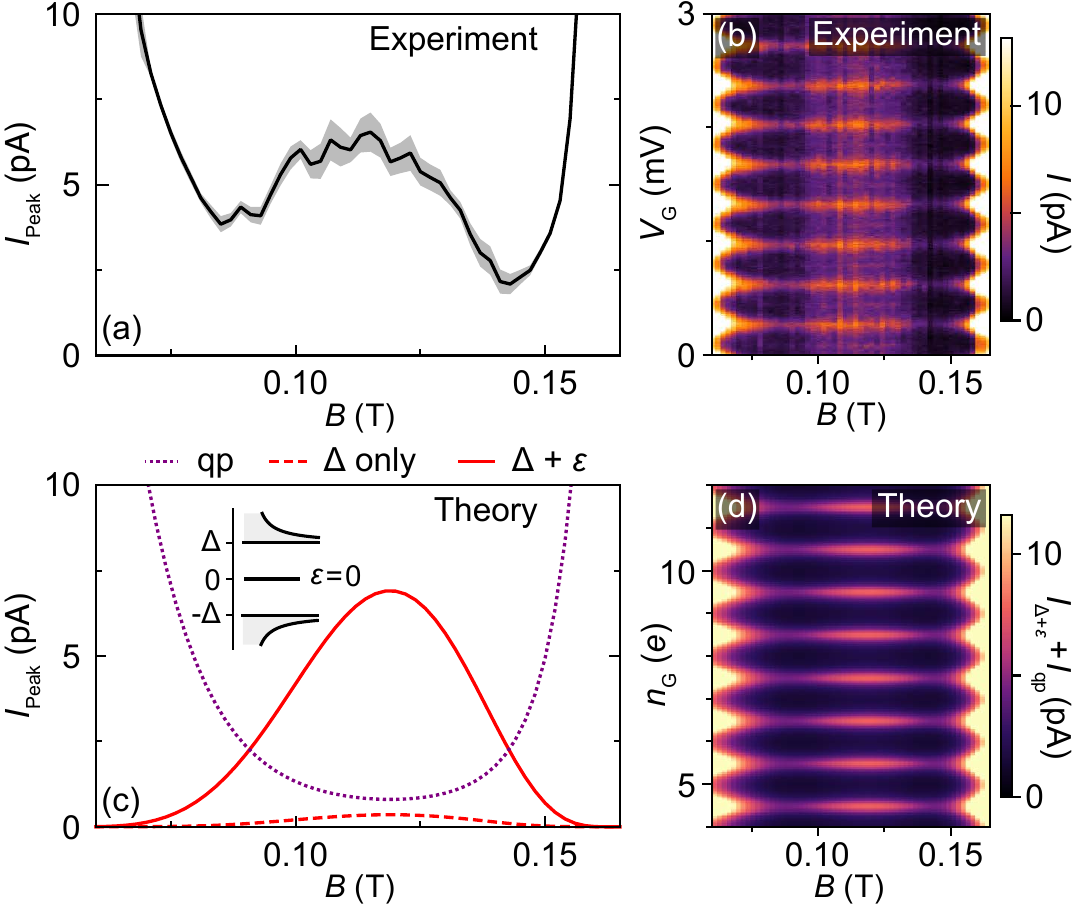}
\caption{\label{fig:5}
(a) Average Coulomb-peak current, $I_{\rm Peak}$, from the data in Fig.~\ref{fig:3}(a), as a function of axial magnetic field, $B$.
The gray band indicates the uncertainty from standard deviation of the peak heights.
(b) First-lobe current, $I$, generated in response to the ac voltage excitation, as a function of island-gate voltage, V$_{G}$, and $B$.
(c) Theoretical current through the device as a function of $B$.
Dotted purple curve is the quasiparticle current given by Eq.~\eqref{eq:2}.
Dashed red curve is the calculated supercurrent contribution only from the states in the continuum.
Solid curve is the calculated supercurrent, including the contribution from zero-energy subgap states in both leads and the island. 
(d) Calculated current in the first lobe as a function of charge offset, $n_{\rm G}$, and $B$, including the normal quasiparticles current and the supercurrent contributions from the continuum and zero-energy subgap state.
}
\end{figure}

To compare the model to experiment, we integrate the average Coulomb-peak conductance over the ac~voltage excitation amplitude (2~$\mu$V) to obtain the measured current, $I_{\rm Peak}$ (solid black curve in Fig.~\ref{fig:4}).
We estimate  the normal quasiparticle current using
\begin{equation}\label{eq:2}
    I_{\rm qp} = I_{\rm N}\,\exp{[-\Delta(B)/k_{\rm B}T]},
\end{equation}
with independently measured electron temperature [Fig.~\ref{fig:2}(d)], field-dependent gap, $\Delta(B)$, (Fig.~\ref{fig:4}, inset), and the normal-state current, $I_{\rm N}$, taken from the destructive regime (dotted purple curve in Fig.~\ref{fig:4}).
Note that away from the destructive regime, $I_{\rm qp}$ is suppressed exponentially and the experimental $I_{\rm Peak}$ can be taken as a proxy of supercurrent flowing through the island.

The supercurrent from the states in the continuum was calculated  by fitting the tunnel amplitude and using independently obtained values for effective temperature and tunnel-rate ratio [Fig.~\ref{fig:2}(d)], $\Delta(B)$, and zero-field $I_{\rm Peak}$ (see Supplemental Material~\cite{Supplement} and Refs.~\cite{Cuevas1996, Scheer1997} therein for details).
Absent the inclusion of subgap states, the model agrees with the experiment in the zeroth lobe but fails to capture the observed bump in $I_{\rm Peak}$ in the middle of the first lobe (compare solid black and dashed red curves in Fig.~\ref{fig:4}).

We next examine the simplest extension of the model by including a single subgap state at $\varepsilon=0$ in the two leads and the island. We account for the change in localization length by linearly scaling the bound-state coupling rates with $\Delta$.
The resulting supercurrent effectively reproduces the experimental data, providing an explanation for the current enhancement observed in the middle of the first lobe [Figs.~\ref{fig:5}(a) and \ref{fig:5}(c)].
The combined supercurrent model and quasiparticle current variation with charge offset on the island qualitatively captures the gate dependence of the measured Coulomb blockade in the first lobe  [Figs.~\ref{fig:5}(b) and \ref{fig:5}(d)].
Within the model, the observed even-odd peak spacing [Fig.~\ref{fig:3}(b)] can be interpreted as resulting from a nonzero-energy subgap state in the island.
The uniform Coulomb peak heights suggest that the subgap state consists of equal electron and hole components~\cite{Vaitiekenas2022, Hansen2018}.
For $\varepsilon = 0$, the Cooper-pair splitting process that is typically virtual during supercurrent transport becomes real.
Increasing $\varepsilon$ in the leads does not affect the peak spacing; instead, it suppresses the supercurrent, suggesting that the supercurrent enhancement observed in the middle of the first lobe requires low-energy subgap states in all three wire segments~\cite{Supplement}.
We note that our phenomenological model is insensitive to the origin of the subgap states.
Nevertheless, our findings are fully consistent with previous studies reported in Refs.~\cite{Vaitiekenas2020, Razmadze2020}.%

In summary, we have studied the supercurrent transport through a Coulomb island embedded in semiconducting (InAs) nanowires with a superconducting (Al) full shell.
Applying an axial magnetic field to the wire resulted in the destructive Little-Parks effect with a reentrant supercurrent around one flux quantum.
In the weak coupling regime, the island displayed 2$e$-periodic Coulomb blockade peaks close to zero field.
Around one flux quantum, the peaks were roughly 1$e$ periodic and showed a local enhancement in conductance proportional to the supercurrent through the system.  
To explain the observed behavior, we used a zero-bandwidth model and showed that the supercurrent enhancement can only be explained by including low-energy subgap states in both leads and the island.
These states enable a supercurrent transport mechanism that includes Cooper-pairs splitting, allowing for resonant tunneling of single electrons.

We thank Claus S\o rensen for contributions to materials growth and Shivendra Upadhyay for assistance with nanofabrication.
We acknowledge support from research grants (Project No. 43951 and No. 53097) from VILLUM FONDEN, the Danish National Research Foundation, the Spanish CM “Talento Program” (project No. 2022-T1/IND-24070), European Research Council (Grants Agreement No. 716655 and No. 856526), and NanoLund.

\bibliography{bibfile}

\end{document}


\title{Supplemental Material:\\Supercurrent transport through 1$e$-periodic full-shell Coulomb islands}

\author{D.~Razmadze}
\altaffiliation{These authors contributed equally to this work.}
\affiliation{Center for Quantum Devices, Niels Bohr Institute, University of Copenhagen, 2100 Copenhagen, Denmark}
\author{R.~Seoane~Souto}
\altaffiliation{These authors contributed equally to this work.}
\affiliation{Center for Quantum Devices, Niels Bohr Institute, University of Copenhagen, 2100 Copenhagen, Denmark}
\affiliation{Division of Solid State Physics and NanoLund, Lund University, 22100 Lund, Sweden}
\affiliation{Instituto de Ciencia de Materiales de Madrid (ICMM), Consejo Superior de Investigaciones Científicas (CSIC),
Sor Juana Inés de la Cruz 3, 28049 Madrid, Spain}
\author{E.~C.~T.~O'Farrell}
\affiliation{Center for Quantum Devices, Niels Bohr Institute, University of Copenhagen, 2100 Copenhagen, Denmark}
\author{P.~Krogstrup}
\affiliation{Center for Quantum Devices, Niels Bohr Institute, University of Copenhagen, 2100 Copenhagen, Denmark}
\author{M.~Leijnse}
\affiliation{Center for Quantum Devices, Niels Bohr Institute, University of Copenhagen, 2100 Copenhagen, Denmark}
\affiliation{Division of Solid State Physics and NanoLund, Lund University, 22100 Lund, Sweden}
\author{C.~M.~Marcus}
\affiliation{Center for Quantum Devices, Niels Bohr Institute, University of Copenhagen, 2100 Copenhagen, Denmark}
\author{S.~Vaitiek\.{e}nas}
\affiliation{Center for Quantum Devices, Niels Bohr Institute, University of Copenhagen, 2100 Copenhagen, Denmark}

\date{\today}

\maketitle

\renewcommand\thefigure{S\arabic{figure}}
\renewcommand{\tablename}{Table.~S}
\renewcommand{\thetable}{\arabic{table}}

\section{Sample preparation}
InAs nanowires with hexagonal cores were grown to a length of $\sim 10~\mu$m and diameter of $\sim 120$~nm using molecular beam epitaxy~\cite{Krogstrup2015}.
An epitaxial Al shell with a thickness of $\sim 30~$nm was grown \textit{in situ} on all six wire facets.  
Devices were fabricated using standard electron beam lithography on highly n-doped Si substrate with 200~nm SiOx capping.
Coulomb islands were defined by selective Al wet etch using MF-321 photoresist developer with 60~s etch time at room temperature.
An adhesion promoter (AR 300-80 new) was used to improve the etch resolution.
The wire ends were contacted with Al (25~nm) leads after Ar-ion milling inside the metal evaporation chamber  (4.5 min, 1 mTorr).
Devices were then coated with a HfOx (8~nm) dielectric layer, followed by Ti/Au (5/200~nm) gate-electrode deposition.

\section{Measurements}
The samples were studied in a dilution refrigerator with a base temperature of 20~mK.
A three-axis (1,\,1,\,6)~T vector magnet was used to align and apply the magnetic field along the wires.
Transport measurements were performed using standard ac lock-in techniques at 86~Hz. 
Four-terminal differential resistance, $dV/dI$, was measured using an ac-current excitation of 0.2~nA. 
Two-terminal differential conductance, $dI/dV$, measurements were performed with 2~$\mu$V ac-voltage excitation.

\begin{figure}[b!]
\includegraphics[width=\linewidth]{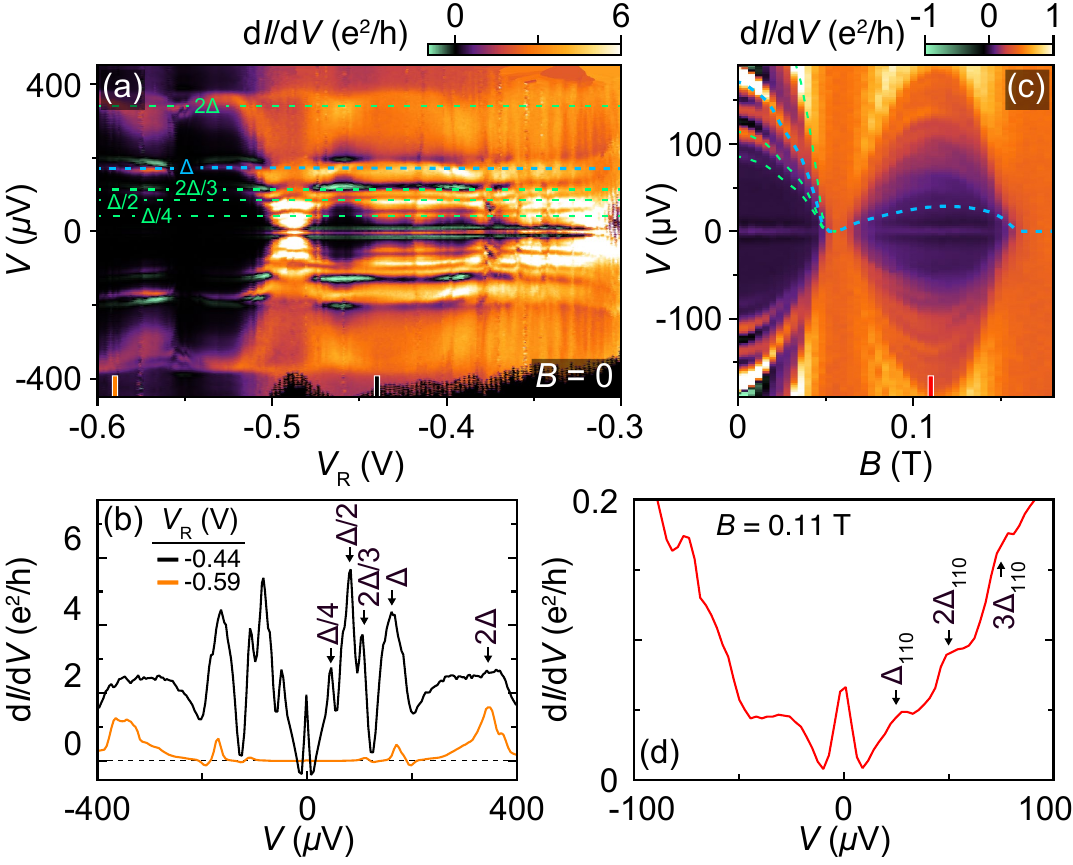}
\caption{\label{fig:S1}
(a) Differential conductance, $dI/dV$, as a function of voltage bias, $V$, and right-junction gate voltage, $V_{\rm R}$, measured at $V_{\rm L}=1$~V.
The spectrum features a series of multiple Andreev reflection peaks (dashed lines), indicating the superconducting gap of $\Delta = 170~\mu$eV.
(b) Linecuts taken from (a) display conductance spectra with multiple Andreev reflection peaks in open (black) and tunneling (orange) regimes.
(c) Evolution of $dI/dV$ with $V$ and the axial magnetic field, $B$, showing the gap closing in the destructive regime and reopening in the first lobe.
(d) Linecut taken from~(c) in the middle of the first lobe, at $B=110$~mT, shows tunneling-conductance spectrum with a spectral gap at $\Delta_{110} \sim 25~\mu$eV and additional equidistantly spaced states at $2\Delta_{110}$ and $3\Delta_{110}$.
}
\end{figure}

\section{Gap spectroscopy}
To study the superconducting gap, $\Delta$, evolution with the axial magnetic field, $B$, the barrier in one of the junctions was lowered by applying a positive gate voltage and tuning the other junction into the tunneling regime.
Example measurement of the differential-conductance spectrum as a function of the right-junction gate voltage, $V_{\rm R}$, measured at $V_{\rm L} = 1$~V, shows several prominent gate-independent features extending from open to the tunneling regime~[Fig.~\ref{fig:S1}(a)].
We attribute these features to multiple Andreev reflections (MARs)~\cite{Cuevas1996, Scheer1997} taking place at $V=2\Delta/e\,n$ (where $n$ is an integer), yielding zero-field gap $\Delta = 170~\mu$eV [Fig.~\ref{fig:S1}(b)].
Field-dependence of the spectrum measured with the left junction in the tunneling regime ($V_{\rm L}$ = -0.151~V, $V_{\rm R}$ = 2~V) shows a gap closing in the destructive regime and reopening in the first lobe, see Fig.~\ref{fig:S1}(c), blue-dashed line.
In the zeroth lobe, the features associated with MARs scale with $\Delta(B)$.
In the first lobe, the spectrum displays a larger number of states with equidistant energy spacing [Fig.~\ref{fig:S1}(d)].
We interpret these states as Caroli-de Genne-Matricon analogs~\cite{Vaitiekenas2020, SanJose2023} and take the state at the lowest energy as the edge of the spectral gap.

\section{Additional Coulomb Spectroscopy}
\begin{figure}[t!]
\includegraphics[width=\linewidth]{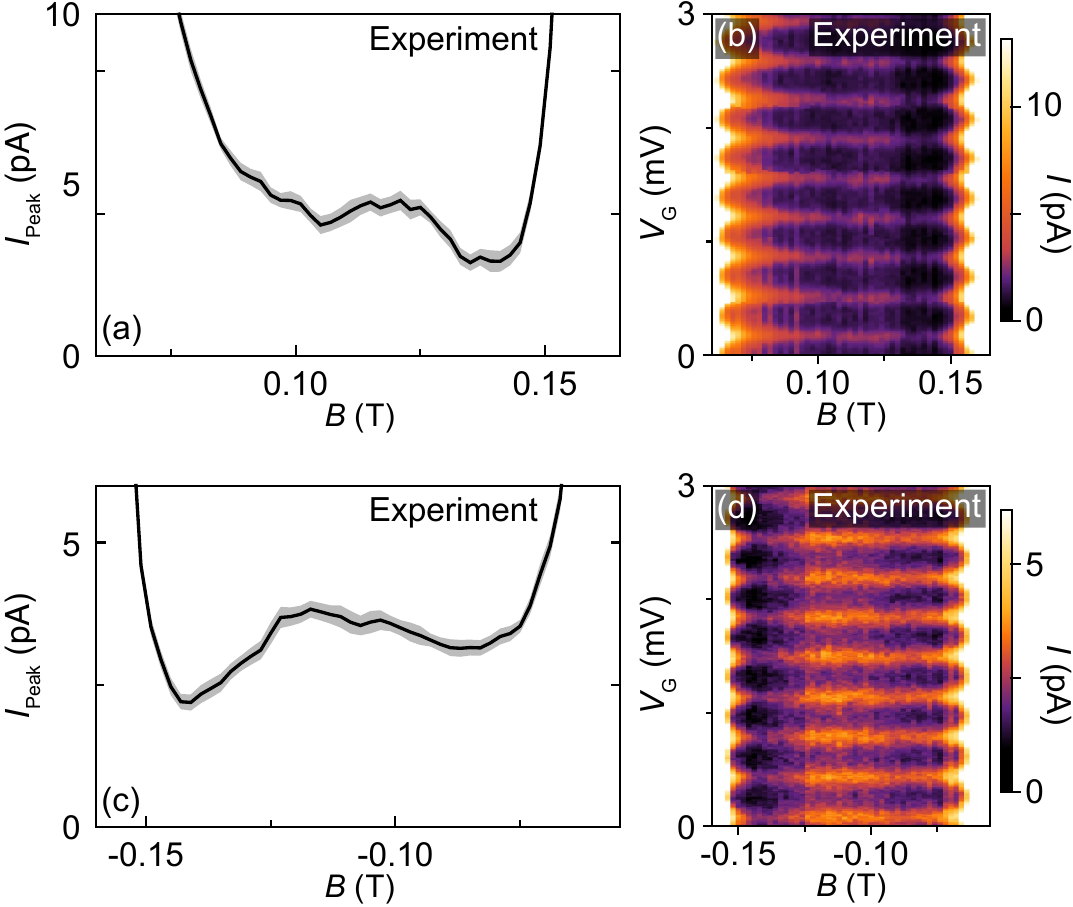}
\caption{\label{fig:S2}
(a) Average Coulomb-peak current, $I_{\rm Peak}$, as a function of axial magnetic field, $B$, measured for device 1 at $V_{\rm L}$ = -0.165~V and $V_{\rm R}$ = -0.394~V.
The gray band indicates the uncertainty from standard deviation of the peak heights.
(b) First-lobe current, $I$, associated with the data in (a), as a function of island-gate voltage, V$_{G}$, and $B$.
(c) and (d) Same as (a) and (b) but measured at $V_{\rm L}$ = -0.17~V and $V_{\rm R}$ = -0.30~V.
}
\end{figure}

\begin{figure}[t!]
\includegraphics[width=\linewidth]{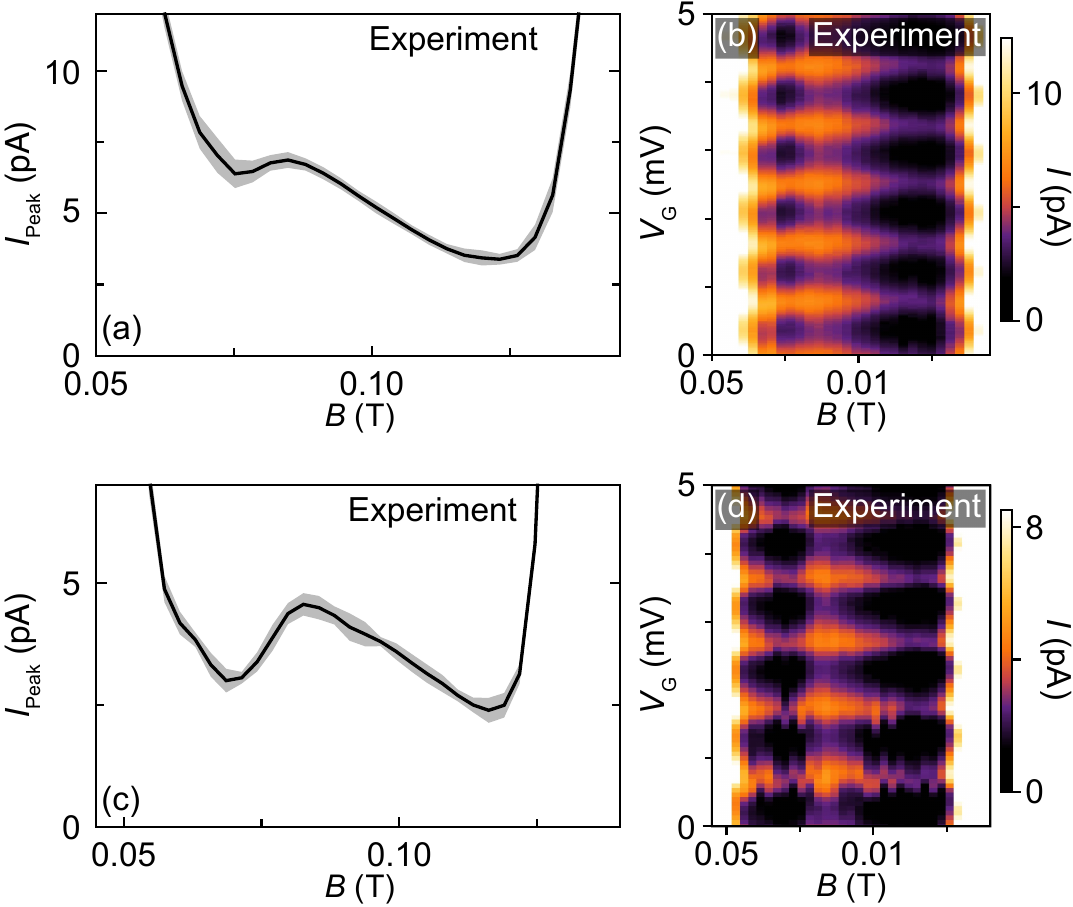}
\caption{\label{fig:S3}
(a) Average Coulomb-peak current, $I_{\rm Peak}$, as a function of axial magnetic field, $B$, measured for device 2 (lithographically similar to device 1) at $V_{\rm L}$ = -0.30~V and $V_{\rm R}$ = -0.35~V.
The gray band indicates the uncertainty from standard deviation of the peak heights.
(b) First-lobe current, $I$, associated with the data in (a), as a function of island-gate voltage, V$_{G}$, and $B$.
(c) and (d) Same as (a) and (b) but measured for device 3 (lithographically similar to device 1) at $V_{\rm L}$ = -0.42~V and $V_{\rm R}$ = -0.20~V
}
\end{figure}

Additional Coulomb spectroscopy data measured for device 1 in different junction-gate voltage configurations display qualitatively the same behavior as the data presented in the main text (Fig.~\ref{fig:S2}). Two additional devices~2 and 3, each comprising a 700~nm island, show similar results (Fig.~\ref{fig:S3}).
In both Figs.~\ref{fig:S2} and \ref{fig:S3} we show examples of the first-lobe current, $I$, obtained by integrating differential conductance maps over the ac voltage excitation window (2~$\mu$V), as a function of island-gate voltage, $V_{\rm G}$, and axial magnetic field, $B$, along with the plots of the extracted $B$-dependence of the average Coulomb-peak current, $I_{\rm Peak}$.
In all cases, a local maximum of $I_{\rm Peak}$ is observed around one applied flux quantum.

\begin{figure}[t!]
\includegraphics[width=\linewidth]{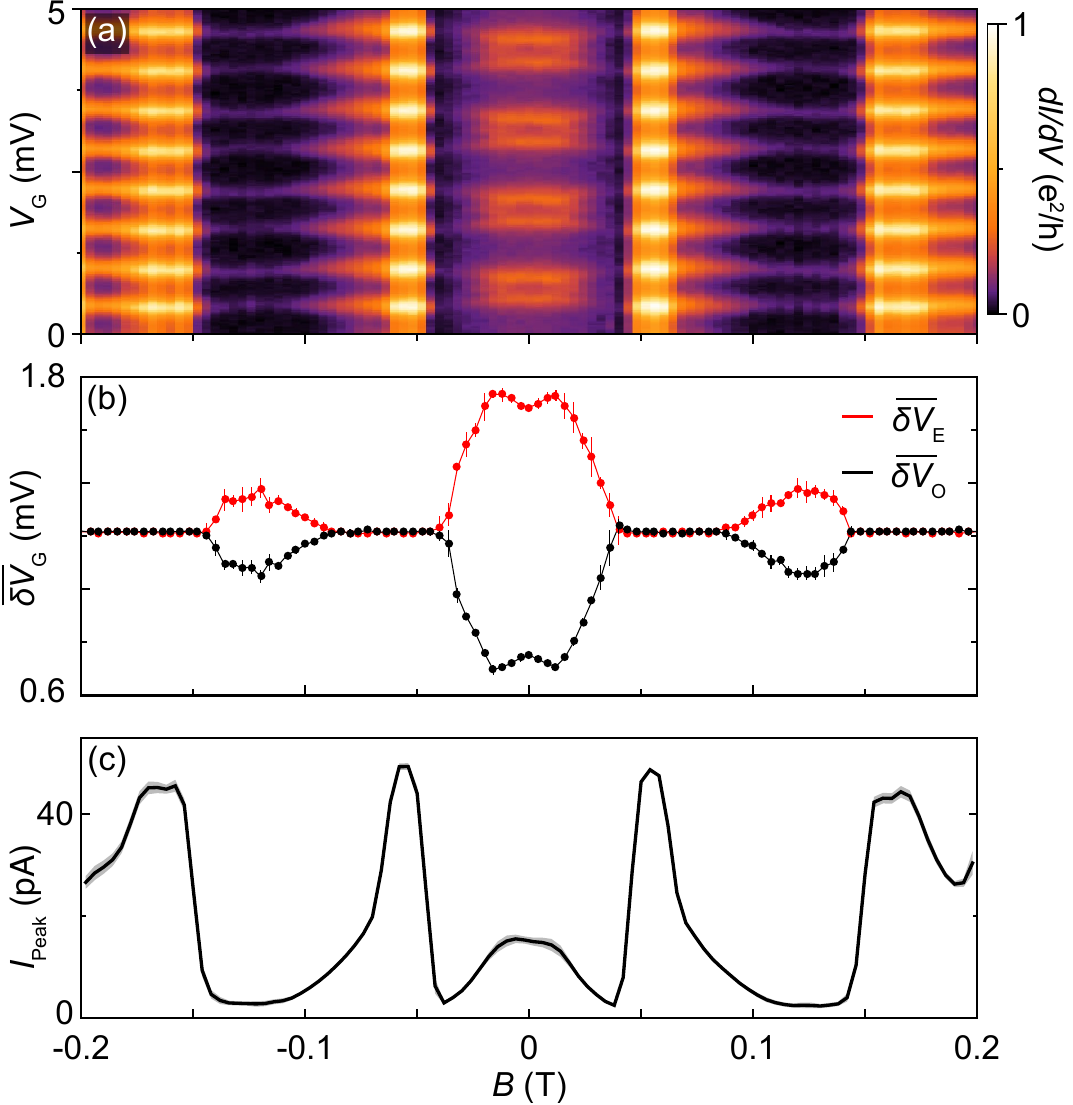}
\caption{\label{fig:S4}
(a) Zero-bias differential conductance, $dI/dV$, measured for device 4 (lithographically similar to device 1, except with a shorter island of 300~nm) in the weak coupling regime ($V_{\rm L}$ = -0.160~V and $V_{\rm R}$ = -0.127~V), showing Coulomb blockade evolution as a function of island-gate voltage, $V_{\rm G}$, and axial magnetic field, $B$.
(b) Average peak spacing, $\overline{\delta V}$, for even (red) and odd (black) Coulomb valleys, extracted from the data in (a).
(c) Average Coulomb-peak current, $I_{\rm Peak}$, as a function of axial magnetic field, $B$, extracted by integrating the differential conductance in (a) over the ac voltage excitation window.
The gray band indicates the uncertainty from standard deviation of the peak heights.
}
\end{figure}

Device 4, with a shorter (300~nm) island, exhibits a qualitatively different behavior (Fig.~\ref{fig:S4}).
Coulomb peaks display even-odd spacing already at $B=0$, suggesting $E_{\rm C} > \Delta$, see Fig.~\ref{fig:S4}(a).
In the first lobe, the peak spacings show a strong even-odd alternation, with the maximal even-odd peak spacing difference corresponding to $\varepsilon \sim 20~\mu$eV [Fig.~\ref{fig:S4}(b)], comparable to the size of the superconducting gap in the first lobe, see Fig.~\ref{fig:S1}(c).
The corresponding average Coulomb-peak current does not feature a local enhancement in the first lobe, see Fig.~\ref{fig:S4}(c).
We note that these results are in agreement with the supercurrent model, suggesting that a low energy sub-gap state in the island ($\varepsilon_{\rm I} \sim 0$) is needed to observe the supercurrent enhancement in the first lobe.

\section{Supercurrent model}
We describe the nanowire device by the following Hamiltonian
\begin{equation}
    H=H_{\rm L}+H_{\rm I}+H_{\rm R}+H_{\rm C}\,,
\end{equation}
where the indices $L$, $I$, and $R$ refer to the left lead, the central island, and the right lead of the device, and $C$ refers to the coupling between the three sections.
To account for the contribution from the states above the superconducting gap, we describe the continuum of states as a discrete state at energy $\Delta_i$ (with $i \in \{L,I,R\}$) using  the zero-bandwidth approximation~\cite{Affleck2000, Vecino2003, Grove2018}.
In addition, we assume that each section can host a subgap state at energy $\varepsilon_{i}$.
With this in mind, the Hamiltonian for the leads can be written as
\begin{equation}\label{eq:3}
    H_{\nu}=\Delta_{\nu}\sum_\sigma \zeta_{\nu\sigma}^{\dagger}\zeta_{\nu\sigma}+\varepsilon_{\nu} \gamma_{\nu}^{\dagger}\gamma_{\nu}\,,
\end{equation}
where $\zeta_{\nu\sigma}=u_{\nu c} c^{\dagger}_{\nu\sigma}+s_\sigma v_{\nu c} e^{-i\phi_\nu} c^{\dagger}_{\nu\sigma}$ is the Bogoliubov-de Gennes (BdG) operator for the quasiparticle states at energy $\Delta$ in the lead $\nu \in \{L, R\}$ and spin $\sigma \in \{\uparrow, \downarrow\}$, with $c_{\nu \sigma}$ being the annihilation operator of an electron and $s_{\uparrow/\downarrow}=\pm1$ depending on the spin.
The BdG coefficients for the states in continuum are $u_{\nu c}=v_{\nu c}=1/\sqrt{2}$.
We use a gauge where the superconducting phase, $\phi_\nu$, appears only in the $v$ coefficients and take the gap $\Delta_\nu$ to be real.
We also define the phase difference between the leads as $\phi=\phi_{\rm L}-\phi_{\rm R}$.
We describe a spin-polarized bound state through the BdG operator  $\gamma_{\nu}=u_{\nu s} d_\nu+v_{\nu s} e^{-i\phi_\nu} d^{\dagger}_{\nu}$, where $u_{\nu s}$ and $v_{\nu s}$ are the BdG coefficients for the boundstates and $d_\nu$ is the annihilation operator of an electron at $\varepsilon_\nu$.

The island is described by
\begin{equation}
    H_{\rm I}=\Delta_{\rm I}\sum_\sigma \zeta_{\rm I\sigma}^{\dagger}\zeta_{\rm I\sigma}+\varepsilon_{\rm I}\gamma_{\rm I}^{\dagger} \gamma_{\rm I}+E_{\rm C} (n-n_{\rm G})^2\,,
\end{equation}
where $E_{\rm C}$ is the charging energy, $n$ is the electron number operator, and $n_{\rm G}$ is the charge offset of the island. As before, the BdG operator for the states at energy $\Delta_{\rm I}$ is given by
$\zeta_{\rm I\sigma}=u_{{\rm I} c} c_{\rm I\sigma}+s_\sigma v_{{\rm I} c} e^{-i\phi_{\rm I}} c^{\dagger}_{\rm I\sigma}$, where $\phi_{\rm I}$ is now an operator and $e^{-i\phi_{\rm I}}$ destroys a Cooper pair in the island.
The BdG operator for the island subgap state is given by
\begin{equation}
    \gamma_{\rm I}=u_{{\rm I}s} d_{\rm I}+v_{{\rm I}s} d_{\rm I}^{\dagger} e^{-i\phi_I}\,.
\end{equation}
For simplicity, we consider $\Delta_\nu=\Delta_{\rm I}\equiv\Delta$ hereafter. The state of the system is characterized by $\left|\vec{m},n\right\rangle$, where $\vec{m}$ is the occupation of the quasiparticle states above the Fermi level (three in each segment) and $n$ is the total number of electrons on the island, including Cooper pairs.

Next, we describe the coupling between the states in the three different regions.
The exchange of electrons between the subgap states in the three regions is given by
\begin{equation}
    H^{s-s}_{\rm C}=\sum_\nu t^{s-s}_{\nu}d^{\dagger}_{\nu} d_{\rm I}+\mbox{H.c.}\,,
\end{equation}
which contains information about the normal tunneling (terms proportional to $u_{\nu s} u_{{\rm I}s}$) and the Andreev reflection (remaining terms) changing the number of Cooper pairs on the island or in the leads.
Here, $t^{s-s}$ is the tunneling amplitude, taken to be real.

\begin{figure}[t!]
\includegraphics[width=\linewidth]{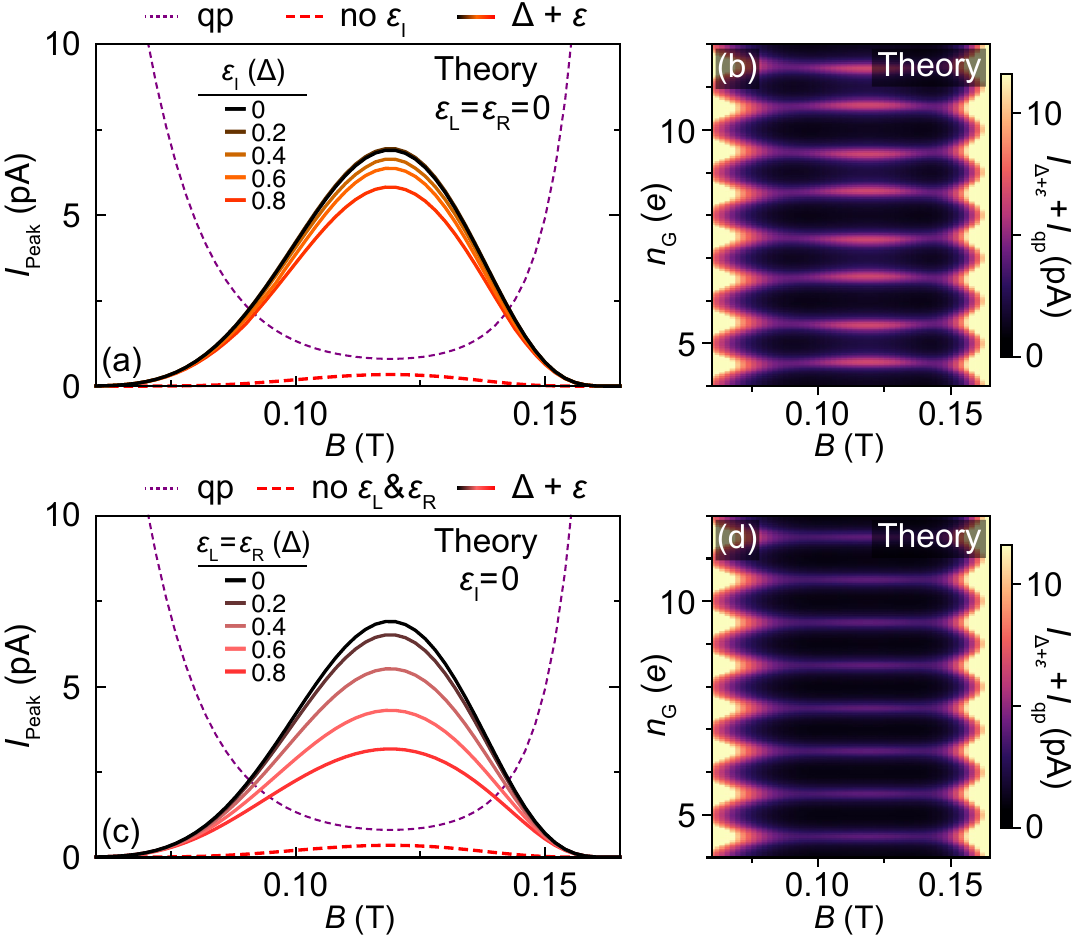}
\caption{\label{fig:S5}
(a) Calculated current as a function of axial-magnetic field in case of zero-energy subgap states in the leads ($\varepsilon_{\rm L} = \varepsilon_{\rm R} = 0$) and varying energy subgap state in the island, $\varepsilon_{\rm I}$.
Dotted purple curve is Eq.~(2) from the main text, with effective electron temperature $T=80$~mK and tunnel couplings $\hbar\Gamma_{\rm L}=\hbar\Gamma_{\rm R}/7=10$~$\mu$eV, showing the normal quasiparticle current.
Dashed red curve is the calculated supercurrent contribution from the states in the continuum and the zero-energy states in the leads, but no subgap state in the island.
Solid curves are the calculated supercurrent, including the contribution from the subgap state in the island at various energies ranging from $\varepsilon_{\rm I} = 0$ to $0.8\Delta$. 
(b) Theoretical calculation of the total current in the first lobe as a function of charge offset, $n_{\rm G}$, and $B$, including the normal quasiparticles current and the supercurrent contributions from the continuum, zero-energy subgap state in the leads, and $\varepsilon_{\rm I} = 0.8 \Delta$.
(c) Same as (a) but for a fixed $\varepsilon_{\rm I} = 0$ and varying $\varepsilon_{\rm L} = \varepsilon_{\rm R}$.
(d) Same as (b) but for $\varepsilon_{\rm I} = 0$ and $\varepsilon_{\rm L} = \varepsilon_{\rm R} = 0.8 \Delta$.
The supercurrent calculations were preformed with the BdG coefficients $u = v = 1/\sqrt{2}$ for all the states, $\phi=\pi/2$, and tunnel amplitudes $t^{c-c}_{\rm L} = t^{c-c}_{\rm R}/\sqrt{7}=0.34\Delta$, $t^{s-s}_{\rm L}=t^{s-s}_{\rm R}/\sqrt{7}=0.14\Delta$, and $t^{s-c}_{\nu}=t^{c-s}_{\nu}=0$.
}
\end{figure}

The current through the continuum is described by
\begin{equation}
    H^{c-c}_{\rm C}=\sum_{\sigma=\uparrow,\downarrow}\sum_\nu t^{c-c}_{\nu}c^{\dagger}_{\nu\sigma} c_{{\rm I}\sigma}+\mbox{H.c.}
\end{equation}
Finally, we also consider the possible hybrid terms, describing the tunneling of electrons that involve the states at $\Delta_i$ and $\varepsilon_i$ in two given regions,
\begin{equation}
\begin{split}
    H^{c-s}_{\rm C}&=\sum_\nu t^{c-s}_{\nu}c^{\dagger}_{\nu\uparrow} d_{\rm I}+\mbox{H.c.}\,,\\
    H^{s-c}_{\rm C}&=\sum_\nu t^{s-c}_{\nu}d^{\dagger}_{\nu} c_{\rm I\uparrow}+\mbox{H.c.}\,,
\end{split}
\end{equation}
where the quantization axis has been chosen to match the spin direction of the subgap state.
We diagonalize the state of the system exactly, obtaining the eigenergies $E_k(\phi)$ that describe the free energy through
\begin{equation}
    F(\phi)=-T\log\sum_k e^{E_k(\phi)/k_B T}\,,
\end{equation}
where $T$ is the temperature of the system.
To reduce the computational cost of the calculations, we truncate the variance of the number of Cooper pairs in the island, ensuring convergence.
With this, the supercurrent is given by
\begin{equation}
    I(\phi)=\frac{2e}{\hbar}\frac{\partial F(\phi)}{\partial \phi}\,,
\end{equation}
where the critical current is given by $I_c=\max_{\phi} |I(\phi)|$.

\begin{figure}[t!]
\includegraphics[width=\linewidth]{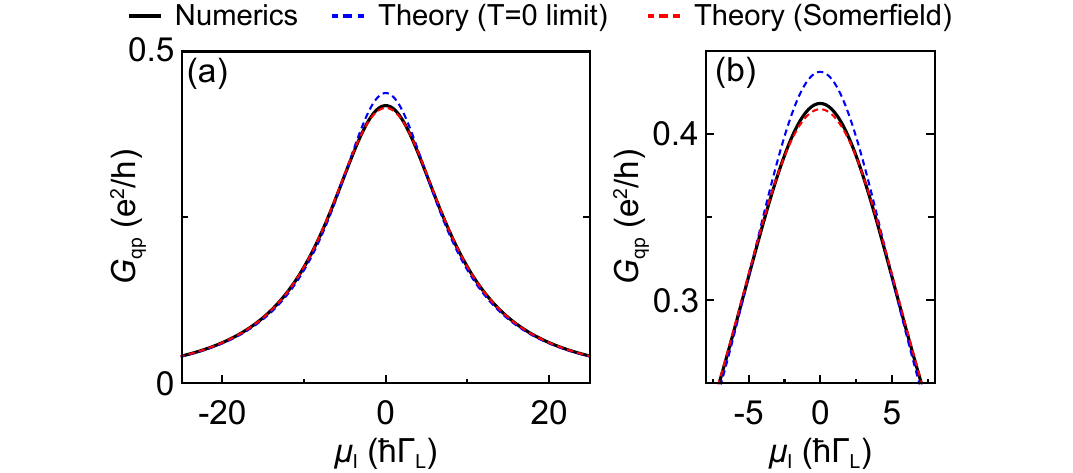}
\caption{\label{fig:S6}
(a) and (b) Different ranges of the calculated normal state conductance, $G_{\rm qp}$, as a function of the chemical potential of the island, $\mu_{\rm I}$.
The zero temperature limit [blue-dashed curve given by Eq.~\eqref{eq:5}] and the Sommerfeld expansion [red-dashed curve given by Eq.~\eqref{eq:6}] are benchmarked with the numerical result obtained by differentiating Eq. \eqref{eq:4} (black curve).
The calculations were done using $\Gamma_L=7\,\Gamma_{\rm R}$, $k_{\rm B} T=\hbar\Gamma_{\rm L}$, and $V=0$.
}
\end{figure}

The results of calculated supercurrent without subgap states ($\varepsilon_i \gg \Delta_i$) are shown in the main-text Fig.~4, where the only fit parameters are the tunnel amplitudes, $t^{c-c}_{\nu}$, while the remaining parameters are taken from independent measurements (see the main text).
An example case including a zero-energy subgap state in all three segments ($\varepsilon_{\rm L}$ = $\varepsilon_{\rm I}$ = $\varepsilon_{\rm R}$ = 0) is summarized in the main-text Fig.~5.

To better understand how the low-energy excitations affect the supercurrent, we systematically study Coulomb spectra varying the energies of the subgap states in the three segments (Fig.~\ref{fig:S5}).
Increasing $\varepsilon_{\rm I}$ away from zero (while keeping $\varepsilon_{\rm L} = \varepsilon_{\rm R} = 0$) results in an even-odd peak spacing, see Figs.~\ref{fig:S5}(a) and \ref{fig:S5}(b).
In contrast, varying $\varepsilon_{\rm L} = \varepsilon_{\rm R}$ (while keeping $\varepsilon_{\rm I} = 0$) reduces the supercurrent enhancement in the first lobe, but does not affect the peak spacing, see Figs.~\ref{fig:S5}(c) and \ref{fig:S5}(d).

We emphasize that a model without subgap states cannot reproduce the current enhancement in the first lobe. Similarly, the model containing a subgap state only in the island cannot reproduce the observed supercurrent enhancement and the homogeneous Coulomb-peak height evolution.
This leads to the conclusion that bound states in the leads are needed to reproduce the experimental observations.

\section{Normal conductance}

\subsection{Destructive regime}
In the destructive regime, the system can be described as a metallic island with charging energy $E_C$.
In this regime, the normal current through the system is given by~\cite{Thijssen2008}
\begin{equation}\label{eq:4}
    I_{\rm N}=-\frac{e}{h}\int D(\omega)\frac{\Gamma_{\rm L} \Gamma_{\rm R}}{\Gamma_{\Sigma}}\left[f_{\rm L}(\omega-\mu_{\rm L})-f_{\rm R}(\omega-\mu_{\rm R})\right]d\omega
\end{equation}
where $f_\nu$ is the Fermi distribution function, $\mu_\nu$ is the chemical potential of a lead defined by the voltage bias $eV = \mu_{\rm L}-\mu_{\rm R}$, $\Gamma_\nu$ is the island-lead tunneling rate, and $\Gamma_{\Sigma}=\Gamma_{\rm L}+\Gamma_{\rm R}$.
Here,
\begin{equation}
    D(\omega)=\sum_k\frac{\hbar\Gamma_{\Sigma}}{(\omega-\omega_k)^2+\hbar^2\Gamma_{\Sigma}^2}\,,
\end{equation}
where the sum runs over the island states.
In this derivation, we consider that the island is metallic, and the energy difference when adding one electron is determined by the charging energy, $\omega_{k+1}-\omega_k=E_{\rm C}$.

At zero temperature and close to zero applied bias ($\mu_L=\mu_R\equiv\mu$), the conductance through the system is given by
\begin{equation}\label{eq:5}
    \left.G_{\rm N}\right|_{T\to0}=\frac{dI_{\rm N}}{dV}=4\frac{e^2}{h}\frac{\Gamma_{\rm L}\Gamma_{\rm R}}{\Gamma_{\Sigma}}D(\mu)\,.
\end{equation}
At the lowest order in temperature, the correction to the conductance can be calculated using the Sommerfeld expansion, giving
\begin{equation}\label{eq:6}
    G_{\rm N}=\left.G_{\rm N}\right|_{T\to0}+\frac{2\pi\left(e\,k_B T\right)^2}{3h}\frac{\Gamma_{\rm L}\Gamma_{\rm R}}{\Gamma_{\Sigma}}\frac{\partial^2 D(\mu)}{\partial \mu^2}\,.
\end{equation}

Figure~\ref{fig:S6} shows a comparison between the numerically obtained normal-state conductance [differentiated Eq.~\eqref{eq:4}] around a Coulomb resonance and the theoretical zero-temperature conductance [Eq.~\eqref{eq:5}]  as well as conductance with the Sommerfeld temperature expansion [Eq.~\eqref{eq:6}].
For $k_{\rm B} T\lesssim \Gamma_{\Sigma}$, the first-order in temperature correction shows a better agreement with the numerical results.

\subsection{Superconductivity effect}
The superconducting pairing opens a gap in the quasiparticle spectrum with an external-field dependent amplitude $\Delta(B)$. At $V=0$, the quasiparticle conductance is given by
\begin{equation}
    G_{\rm qp}=G_{\rm N} \int_{0}^\infty d\omega f(\omega)D_{\rm S}(\omega)
\end{equation}
where expressions for $G_{\rm N}$ are provided in the previous section. We approximate the continuum density of states using the conventional BCS expression, given by
\begin{equation}
    D_S(\omega)={\rm Im}\frac{\omega+i\eta}{\sqrt{\Delta^2-(\omega+i\eta)^2}}\,,
\end{equation}
where $\eta$ is the so-called Dynes parameter, broadening the density of states at the edge of the gap.
For $\Delta\gg k_B T$, the normal conductance can be approximated as
\begin{equation}
    G_{\rm qp}\approx G_{\rm N} e^{-\Delta/k_B T}\,,
\end{equation}
illustrating that the normal conductance is proportional to the density of quasiparticles excited above the superconducting gap, which decreases exponentially with the gap.

To compare these results to the experiment on the same footing as the supercurrent model (see main-text Fig.~4), we convert the calculated conductance to current, which, in the linear regime, is given by 
\begin{equation}
    \label{eq:7}
    I_{\rm qp}=I_{\rm N} \int_{0}^\infty d\omega f(\omega)D_{\rm S}(\omega)\,.
\end{equation} 
We note that $\Delta(B)$ and the electron temperature $T\sim80$~mK are known from independent measurements [see Fig.~\ref{fig:S1}(c) and main-text Fig.~2(a)], and the only fit parameter is the normal-state quasiparticle current $I_{\rm N}$.

\bibliography{bibfile}